\documentclass{aastex631}

\shorttitle{The secondary maximum of T CrB caused by irradiation of the red giant}
\shortauthors{U. Munari}

\begin{document}

\title{The secondary maximum of T CrB caused by irradiation of the red giant
by a cooling white dwarf}

\author[0000-0001-6805-9664]{Ulisse Munari}
\affiliation{INAF Astronomical Observatory of Padova, 36012 Asiago (VI), Italy}



\begin{abstract}

Both the 1866 and 1946 outbursts of the recurrent symbiotic nova T CrB have
displayed a mysterious secondary maximum peaking in brightness $\sim$5
months past the primary one.  Common to all previous modeling attempts was
the rejection of plain irradiation of the red giant (RG), on the basis that
the secondary maximum of T CrB would have been out of phase with the transit
at superior conjunction of the RG.  Implicit to this line of reasoning is
the assumption of a {\it constant temperature} for the white dwarf (WD)
irradiating the red giant.  I show by radiative modeling that irradiation of
the RG by a {\it cooling} WD nicely reproduces the photometric evolution
of the secondary maximum, both in terms of brightness and color, removes the
phasing offset, and provides a straightforward explanation that will be easy
to test at the next and imminent outburst.
 
\end{abstract}



\section{Introduction} \label{sec:intro}

T CrB belongs to the rare group of symbiotic recurrent novae, of which only
$\sim$four are known in the Galaxy, the most famous being RS Oph.  The last
eruption of T CrB occoured in 1946 and its light curve was a near-perfect
replica of the previous outburst in 1866 \citep{1946PASP...58..359P}.  The
lightcurve of both events (separated by exactly 128 orbital revolutions) is
characterized by the presence of a broad secondary maximum (II-Max
hereafter), which is seen only in T CrB.

\begin{figure}
\includegraphics[width=16cm]{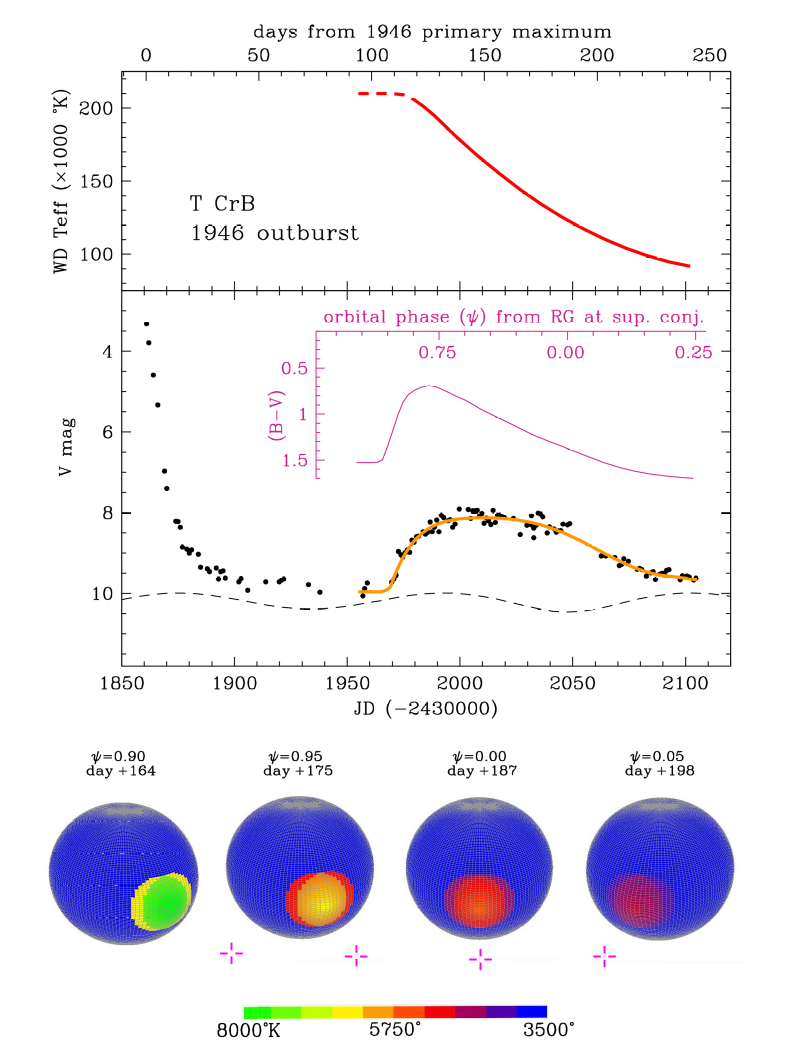}
\caption{The secondary maximum of T CrB is accurately reproduced (orange
line) by irradiating the red giant with the white dwarf companion cooling
according to the temperature profile plotted in the top panel.  The dashed
line shows for reference the ellipsoidal modulation computed for the
Roche-lobe filling red giant.  The inset present the evolution of $($$B$$-$$V$$)$
color.  Days are counted from primary maximum on JD 2431860.854 (= 1946 Feb
9.354 UT), and phase $\psi$ is reckoned from RG passage at superior
conjunction ($\psi$=0.0 on day +187 and JD 2432048).  At the bottom,
examples of the computed radiative models (the crosses highlight the
position of the WD).}
\label{fig1}
\end{figure}

Different explanations (including an accretion episode, irradiation of a
tilted disk, and a second and separate nova eruption) have been proposed for
II-Max
\citep[eg.][]{1976Natur.262..271W,1999ApJ...517L..47H,2023MNRAS.524.3146S}. 
Common to all of them is the rejection of a plain irradiation of the red
giant (RG), on the basis that II-Max is out of phase with the transit at
superior conjunction of RG ($\psi$=0 in Figure~\ref{fig1}).  Implicit to
this line of reasoning is the assumption of a {\it constant temperature} for
the WD irradiating the red giant.

Supported by the results of detailed radiative modeling, I will show that
irradiation of the RG by a {\it cooling} WD nicely fits the photometric
observations of II-Max, both in terms of brightness and color, removes the
phasing problem, and provides a straightforward explanation that will be
easy to test at the next, imminent outburst
\citep{2016NewA...47....7M,2023MNRAS.524.3146S,2023RNAAS...7..145M}.

\section{Radiative Modeling} \label{sec:mod}

A radiative modeling for II-Max has been carried out in physical units,
placing T CrB at the 916~pc distance derived by Gaia and adopting a
$E_{B-V}$=0.05 reddening, a mass of 1.3~M$_\odot$ for the white dwarf
(WD) and 0.93~M$_\odot$ for the red giant (RG), 227.56~days as the orbital
period, $i$=65$^\circ$ for the orbital inclination, and a null eccentricity
\citep{2000AJ....119.1375F}.  The WD is taken to radiate isotropically as a
blackbody, while the surface of the Roche-lobe filling RG is divided into a
256$\times$256 mesh grid, with each area bin radiating according to model
atmospheres taken from \citet{2003IAUS..210P.A20C}, and interpolated to
local $T_{\rm eff}$ and $\log g$.  Coefficients for linear gravity darkening
are derived from \citet{2011A&A...529A..75C}.  A fraction $\eta$ of the
radiation arriving from the WD on the RG is locally absorbed and
re-thermalized, the remaining 1$-$$\eta$ is scattered out as it is.  A
$T_{\rm eff}$=3500$^\circ$K is adopted for the shadowed regions of RG, in
line with its M3III spectral classification.  The binary system is followed
through orbital revolution, and at each step the emitted spectrum is
integrated through the profile of Landolt $B$,$V$ bands and magnitudes
computed, with flux zero-points taken from \citet{1998A&A...333..231B}.

The lightcurve of the 1946 outburst of T CrB is presented in
Figure~\ref{fig1} (dots).  It is built from \citet{1946PASP...58..359P}
observations ported to modern Landolt $V$-band by comparing to APASS
survey \citep{2014CoSka..43..518H} the quoted magnitudes for the eight
original comparison stars.  The spectral evolution of T CrB during the 1946
outburst has been described in detail by eg. 
\citet{1946PASP...58..156S,1947PASP...59...87S,1949ApJ...109...81S} and
\citet{1947ApJ...106..362M}, indicating how the WD was very hot when
II-Max begun on day +109, with coronal lines of [FeX] and [FeXIV] being
persistently strong since day +4, while [FeVII] was still absent.  At the
end of II-Max the WD was still rather hot, albeit cooler, with 
spectra showing [FeX] and [FeVII], but no more [FeXIV].  Also RS Oph
displayed for long a very hot WD during the 2006 and 2021 outbursts, till at
least day +86 as proved by the prominence of [FeX], [FeXIV] in optical
spectra \citep{2022arXiv220301378M} and the strong super-soft emission in
X-rays observations \citep{2022MNRAS.514.1557P}.

During the radiative modeling runs only the WD and the irradiated RG were
considered, the contribution from the nova ejecta to overall brightness
being irrelevant: in fact, the ejecta were already optically thin by day +4
(coronal lines prominent), the M3III spectrum of the RG returned
visible in the blue by day +13, and the classical nebular emission lines
ever developed.  No accretion disk is considered either, by analogy with RS
Oph in which the disk begins reforming only $\sim$120 days past
disappearance of coronal lines \citep{2022RNAAS...6..103M}.  The temperature
of the WD at the beginning of II-Max is assumed to be 220,000~K consistent
with a photoionization origin for [FeXIV], and the radius set to
0.23~R$_\odot$ to fit the constant brightness exhibited by T CrB during the weeks
preceding II-Max.  While the temperature of the WD was let to change, its
radius has been kept fixed through II-Max.

\section{Results} \label{sec:results}

Irradiating the RG by the WD cooling according to the 
temperature profile at the top of Figure~\ref{fig1} returns a perfect match
to the observations of II-Max, as indicated by the overplotted thick orange
line.  Also the computed $($$B$$-$$V$$)$ color (shown in the inset and inclusive
of $E$$($$B$$-$$V$$)$=0.05 interstellar reddening) is in good agreement with
\citet{1946PASP...58..255P} observations which report the bluest color of T
CrB being attained around JD 2431980-90 and being similar to that of a G0
star.  The best fit is reached for $\eta$=0.8, but similarly good fits to
II-Max can be achieved by trading a lower $\eta$ for a higher temperature
and/or radius of the WD.  The temperature of the sub-WD position on the RG
surface (coincident with Lagrange L1 point) relates obviously to WD
temperature: for 0.23~R$_\odot$ WD radius and $\eta$=0.8 it
is 7797, 5972, 4391, and 3578~$^\circ$K for a WD temperature of respectively
200, 150, 100, and 50$\times$10$^3$~$^\circ$K.

{\sc Acknowledgements}.  The support of A.  Frigo and the encouragement by
M.  Giroletti are acknowledged.
 
\bibliography{paper.bib}{}
\bibliographystyle{aasjournal}



\end{document}